\documentclass[aps,prb,superscriptaddress]{revtex4}
\usepackage{subfigure}
\usepackage{amsmath}
\usepackage{amssymb}
\usepackage{graphics}
\usepackage{rotating}
\usepackage{array}
\usepackage{graphicx}
\usepackage{epstopdf}
\usepackage{hyperref}
\usepackage{xcolor}
\newcommand{\ddp}[2]{\frac{\partial #1}{\partial #2}}

\begin{document}

\title{Simulating structural transitions by direct transition current sampling:  the example of LJ$_{38}$}
\author {Massimiliano Picciani}
\affiliation{CEA, DEN, Service de Recherches de M\'etallurgie Physique, F-91191 Gif-sur-Yvette, France}
\author{Manuel Ath\`enes}
\affiliation{CEA, DEN, Service de Recherches de M\'etallurgie Physique, F-91191 Gif-sur-Yvette, France}
\author{Jorge Kurchan}
\affiliation{CNRS; ESPCI, 10 rue Vauquelin, UMR 7636, Paris, France 75005,PMMH}
\author{Julien Tailleur}
\affiliation{School of Physics of Astronomy, SUPA, University of Edinburgh, The King's buildings, Mayfield Road, EH9 3JZ, Edinburgh, UK}

\renewcommand{\baselinestretch}{1.2}\normalsize

\begin{abstract}
Reaction paths and probabilities are inferred, in a usual Monte Carlo
or Molecular Dynamic simulation, directly from the evolution of the
positions of the particles. The process becomes time-consuming in many
interesting cases in which the transition probabilities are small.  A
radically different approach consists of setting up a computation
scheme where the object whose time evolution is simulated is the
transition current itself.  The relevant timescale for such a
computation is the one needed for the transition probability {\em
  rate} to reach a stationary level, and this is usually substantially
shorter than the passage time of an individual system.  As an example,
we show, in the context of the `benchmark' case of 38 particles
interacting via the Lennard-Jones potential (`LJ$_{38}$' cluster), how
this method may be used to explore the reactions that take place
between different phases, recovering efficiently known results and
uncovering new ones with small computational effort.
\end{abstract}

\maketitle 

\section{Introduction}

The evolution of multi-particle systems arising in very diverse
domains ranging from biology to material science is often governed by
activated processes that occur with very low probability, but have a
dramatic effect on the structure. Examples of physical phenomena
controlled by such rare events are protein folding in biology, defect
diffusion and crystal nucleation in condensed matter physics and
cluster rearrangement in chemistry.

Activated events are characterized by the fact that when they occur,
their duration is rather short, e.g. of the order of the picoseconds
in dense molecular systems, but that, in contrast, the typical time
needed for them to start is very large. This separation of time-scale
is the very definition of metastability; its origin may be energetic
(high barriers) or, more likely in high dimensional systems, at least
partly entropic (pathways hard to find). Even though physical times of
activated events can be achieved in computer simulation using
molecular dynamics, monitoring rare fluctuations is computationally
expensive and still remains a challenge because of the long waiting
time for the event to occur.  To overcome this problem, several
different strategies have been developed over the past years.  A
common idea behind these techniques is to bias the dynamics of the
system in order to enhance the occurrence of reaction, or to use
previous knowledge of the outcome of the reaction and to fix the
endpoint of the trajectory. In order to implement this, in many cases
one needs to know an order parameter able to discriminate between the
reactant basin, the saddle regions and the product basin.
 
A set of methods involving the direct sampling of path ensembles has
been developed during the last decade.~\cite{W2003} The strategy is to
restrict paths to the subset of reactive paths, those that interpolate
between reactant and product basins. Examples of such methods are
transition path sampling,~\cite{DBCD1998,BCDG2002} transition
interface sampling.~\cite{VEB2005} Another family of methods such as
metadynamics~\cite{Laio} and forward flux sampling~\cite{VAMFW2007}
simulate the evolution in time of the system, and include some form of
bias that guarantees that the reaction happens, but then of course the
effect of the bias has to be discounted in order to retrieve the true
statistics. Methods differ also in the way rate constants are
estimated,~\cite{AVW2009} and the ability to handle nonequilibrium
steady states.

Observation of rare events and exploration of energy landscapes of a
multi-particle system are indeed two interconnected problems.  The
requirement of knowing the target of the reaction, and in particular a
relevant order parameter that parametrizes the path is problematic in
physical situations where nothing {\em a priori} is known.  Thus, a
preliminary step prior to the calculation of finite-temperature rate
constants using the sampling techniques mentioned above may consist in
locating the saddle points of the energy and the corresponding energy
minima using one of the eigenvector following methods described in the
literature (e.g. the activation-relaxation technique,~\cite{MB1998}
Optim~\cite{MW1999} or the dimer method~\cite{HJ1999}). The technique
in all these methods relies on monitoring the eigenvalues of the
Hessian matrix (the matrix of the second derivatives of the potential
energy), in order to individuate stable or unstable directions,
i.e. minima or saddles. A limitation of these methods is that energy
saddles only correspond directly to the actual barriers for the
dynamics if the temperature is very low, otherwise the entropic
contribution to the dynamics becomes relevant.

Another approach that may be used to explore the {\em free-energy}
(i.e. finite-temperature) barriers between basins
\cite{TK2004,TTK2006} is inspired by Supersymmetric Quantum Mechanics,
which has been used in the context of field theory to derive and
generalize Morse Theory, precisely the analysis of the saddle points
of a function.  Transposing to statistical physics this formalism, one
obtains a family of generalizations of Langevin dynamics, converging
to barriers and reaction paths of different kinds, rather than to the
equilibrium basins.  The resulting method involves the evolution in
phase space of a population of independent trajectories that are
replicated or eliminated according to the value of their Lyapunov
exponent.\cite{TK2004} The theory guarantees that by selecting
trajectories having larger Lyapunov exponents in a specified way, the
bias is just what is needed so that the population describes the
evolution of the {\em transition current}, rather than that of the
configurations,\cite{TK2006} as they would in an unbiased case. The
advantage is that the convergence of the current distribution is much
faster than the typical passage time.

In this paper we shall present a more direct and elementary derivation
of this dynamics without resorting to quantum theory
(section~\ref{theorie}). Extending the more concise demonstration of
Ref.~\onlinecite{TK2003b}, we will show how the modified dynamics
precisely reproduces the evolution in the phase space of the
probability current (or more precisely, the `transition' probability
current) between equilibrium basins, thus achieving a
\textit{probability current sampling} of the system dynamics.  As the
transition current evolves in time, it explores the different
barriers, indicating which states are reached after different passage
times. Starting from an initial equilibrium configuration, far from
equilibrium phenomena are easily sampled, as the simulated current is
an intrinsically out of equilibrium quantity.\cite{TK2004} In section~
\ref{method}, we discuss the actual population dynamic algorithm that
may be used to simulate the probability current. Finally, in order to
assess the performances of this probability current sampling, we
present in section \ref{sec:LJ38} an application to the structural
transitions of a Lennard-Jones cluster, sampled under different
physical conditions. We shall recover known results, and discuss some
new ones, obtained in all cases at rather small computational costs.

\section{Transition currents} \label{theorie}

\subsection{Overdamped Langevin dynamics}
\label{sec:fp}

Let us consider a 3D many-body system composed of \textit{N} particles
in contact with a thermal bath. We denote $r_1,\dots, r_{3N}$ the
configurational degrees of freedom and $V(\{ r_i\}_{i=1\dots 3N})$ the
interaction potential. In the limit of large friction
$\gamma\to\infty$, where inertia can be neglected, the system evolves
with a standard overdamped Langevin dynamics
\begin{equation}
  \label{langov}
 \gamma  \dot{r_{i}} = -\frac{1}{m_i }\frac{\partial V}{\partial
    r_{i}}+\sqrt{\frac{2 \gamma k_{B}T}{m_{i}}}\eta_{i}\;,
\end{equation}
where $\eta_i$ are independent Gaussian white noises of unit variance
and $m_i$ correspond to the particle masses.

The probability to find the system at position ${\bf r}\equiv
(r_1,\cdots r_{3N})$ then evolves with the Fokker-Planck
equation~\cite{R1989}
\begin{equation}
  \label{1.1}
  \frac{\partial P}{\partial
  t}=\sum_{i}\frac{1}{m_i\gamma}\frac{\partial}{\partial r_i}
  \left(k_{B}T\frac \partial{\partial r_i}+\frac{\partial V}{\partial r_i}\right)P\equiv
  -H_{FP}P\;,
\end{equation}
where we have introduced the Fokker-Planck operator $H_{FP}$. We can
use the probability current $J_i\equiv -\frac 1 {\gamma m_i}
\left({k_{B}T}\frac{\partial}{\partial r_i}+\frac{\partial V}{\partial
  r_i}\right)P$ to write Eq.~\ref{1.1} as a continuity equation for
the probability density:
\begin{equation}
  \frac{\partial P}{\partial t}+\sum_{i}\frac{\partial J_i}{\partial r_i}=0\;.
\end{equation}
For systems with separation of time scales, the dynamics can be split
into two regimes. Starting from an arbitrary probability distribution,
$P({\bf r };t)$ relaxes {\em rapidly} into a sum of contributions
centered on the metastable states. At {\em much longer times}, the
rare transitions between the metastable states make $P({\bf r};t)$
relax to the equilibrium distribution.  Two time scales can also be
identified for the dynamics of the probability current. While the
probability density rapidly relaxes into the metastable states, the
probability current converges {\em on the same time scale} to the most
probable transition paths between the metastable states. Then, the
late time relaxation towards equilibrium corresponds to a progressive
vanishing of the current, when forward and backward flux between each
metastable state balance.~\cite{TK2004} { Note that the same line of
  reasoning holds for non-equilibrium systems in which the forces do
  not derive from a global potential. In such systems, the probability
  current never vanishes and converges instead to its steady-state
  value.}

If one were able to simulate the evolution of the probability current,
one would thus have all the knowledge relevant for the transitions
between metastable states, while only having to simulate the system
for relatively short time-scales (similar to the equilibration time
within a metastable state). As mentioned in the introduction,
simulating directly the transition current is the goal of this paper
and we now derive a self-consistent evolution equation for $J_i$.

Let us define the \emph{current operator} $\hat J_i$:
\begin{equation}
  \hat{J}_{i}\equiv \frac 1 {\gamma m_i} \left(k_{B}T
\frac{\partial}{\partial r_i}+ \frac{\partial V}{\partial r_i}\right)\;,
\end{equation}
so that the probability current and the Fokker-Planck operator
becomes
\begin{subequations}
  \begin{align}  
    \label{eqn:newcur}    J_i&=\hat J_i P\\
    \label{eqn:newhfp} H_{FP}&=-\sum_j \frac{\partial}{\partial r_j}\hat{J}_{j}\;.
  \end{align}
\end{subequations}
The evolution of the probability current is then given by
\begin{equation}
  \label{eqn:currevol1}
  \dot J_i=\hat J_i \dot P=-\hat J_i H_{FP} P=-\sum_j \hat J_i \frac{\partial}{\partial r_j} \hat J_j P\;.
\end{equation}
where we have assumed  that $H_{FP}$ does not depend explicitly on time. 
Straightforward algebra shows that $\hat J_i \frac{\partial}{\partial
  r_j}=\frac{\partial}{\partial r_j} \hat J_i
-\frac{\partial^2 V}{\partial r_i \partial r_j}$ and $\hat J_i \hat J_j
= \hat J_j \hat J_i$ which turns  equation \eqref{eqn:currevol1} into
\begin{equation}
  \label{eqn:JofP2}
  \dot J_i=-\sum_j \left(\frac{\partial}{\partial r_j} \hat J_j \hat J_i 
  - \frac{\partial^2 V}{\partial r_i \partial r_j} \hat J_j \right )P\;.
\end{equation}
Using the expressions \eqref{eqn:newcur} and \eqref{eqn:newhfp} for
the currents and Fokker-Planck operator we obtain
\begin{equation}
  \label{eqn:currentevolOD}
  \dot J_i= -H_{FP} J_i - \sum_j \frac{\partial^2 V}{\partial r_i \partial r_j} J_j\;.
\end{equation}
Note that the equations \eqref{eqn:newcur} and \eqref{eqn:JofP2} are
not self-contained: the knowledge of $P({\bf r})$ is required to
compute $J_i$. On the contrary, \eqref{eqn:currentevolOD} depends exclusively on
the current,
 and can readily be used to simulate $J_i$, without
having to compute $P({\bf r})$ beforehand.
The only condition is that the  current distribution at the initial time $J_i^0$ indeed derives from
a probability distribution,  i.e. is of the form:
\begin{equation}
J_i^0\equiv -\frac 1 {\gamma m_i}  \left({k_{B}T}\frac{\partial}{\partial r_i}+\frac{\partial
 V}{\partial r_i}\right)P^0=  -\frac{k_{B}T} {\gamma m_i}e^{-\frac{1}{k_B T} V}\frac{\partial}{\partial r_i}
 \left[e^{\frac{1}{k_B T} V} P^0 \right]
 \label{initial}
 \end{equation}
 This means that the initial current distribution should be such that the quantity $A_i$
 \begin{equation}
A_i =  m_i   e^{\frac{1}{k_B T} V} J_i^0 
\end{equation}
is a gradient, $\frac{\partial A_i}{\partial r_j}=\frac{\partial
  A_j}{\partial r_i}$.  A particularly simple initial condition is
obtained if one assumes that $P^0$ is Gibbsean $P^0 \propto
e^{-\frac{1}{k_B T} V} $ in a region $\Omega$, and zero elsewhere.
Then, from Eq. (\ref{initial}), $J^0_i$ is zero everywhere except on
the surface of $\Omega$, where it takes the form of a vector normal to
the surface of $\Omega$, and with amplitude proportional to the Gibbs
weight.

The evolution of current distribution given by Equation
\eqref{eqn:currentevolOD}, starting from an appropriate initial
current $J^0$ converges to the stationary distribution of currents
between metastable states \emph{on the same time scale} as the usual
Langevin equation converges to metastable-state. It is thus not
necessary to wait for rare events to identify the transition path
between the metastable states, an important improvement over standard
MD methods.  If there are several metastable states and transitions
with different rates, the current distribution at longer times
concentrates on the paths between regions that have not yet mutually
equilibrated, and vanishes in transitions between states that have had
the time to mutually equilibrate.

\subsection{Langevin dynamics with inertia}
\label{sec:kram}
In many physical situations inertia plays an important role and one
cannot rely on overdamped Langevin equations.~\cite{HTB1990} In such
case, the stochastic dynamics of $N$ interacting particles is given by
\begin{subequations}
  \begin{align}
    \dot{r_{i}}&= v_{i} \\ \dot{v_{i}}&=-m_i^{-1} \frac{\partial
      V}{\partial r_{i}}-\gamma \, v_{i}+\sqrt{2 m_i^{-1} \gamma
      k_{B}T}\,\eta_{i}\;,
  \end{align}
  \label{langund}
\end{subequations}
where $(v_1,\cdots,v_n)\equiv {\bf v}$ correspond to the $3N$
velocities.

In order to transpose the derivation of the current evolution given in
 subsection \ref{sec:fp} to the inertial case, we have to consider the
 Kramers evolution equation $\dot{P}({\bf r},{\bf v};t)=-H_{K}P({\bf
 r},{\bf v};t)$ instead of the Fokker-Planck equation:~\cite{TTK2006}
\begin{equation}\label{2.14}
\frac{\partial P}{\partial
t}=\sum_{i}\left[\ddp{}{v_i}\left(\frac{\gamma k_B T}{m_i}
\ddp{}{v_i}+\gamma v_i +\frac 1 {m_i}\ddp{V}{r_i}\right)- \ddp{}{r_i}
v_i\right]P\;.
\end{equation}
As in the overdamped case, this can be written as a conservation
equation where the probability current {\em in phase space} is given
by
\begin{subequations}
  \label{test}
  \begin{align}
    J_{r_i}&= v_i\, P({\bf r},{\bf v};t)\label{eqn:currentr}\\
    J_{v_i}&= -\left(\frac{\gamma k_B T}{m_i}
    \ddp{}{v_i}+\gamma v_i +\frac 1 {m_i}\ddp{V}{r_i}\right) P({\bf r},{\bf v};t)\;.\label{eqn:currentv}
  \end{align}
\end{subequations}
Once again, the current contains all the information about transitions
between metastable states. 
There is however a conceptual difference: the presence of inertia makes
it inherently difficult  (and indeed, useless),  to compute~(\ref{test}) as it stands. 
 The reason is that the {\em phase-space} current 
\begin{equation}
  \label{eqn:hocur}
  J_{r_i}= v_i P_{\rm eq}\;\qquad
  J_{v_i}= -\frac {1}m \ddp{V}{r_i} P_{\rm eq}\;.
\end{equation}
is  non-zero even in canonical equilibrium.
For example, in a harmonic oscillator  $H= \frac{1}{2} \left[ v^2 +  r^2 \right]$ , the phase-space current  in equilibrium   turns clockwise in circles around the origin.

\begin{figure}
 \includegraphics{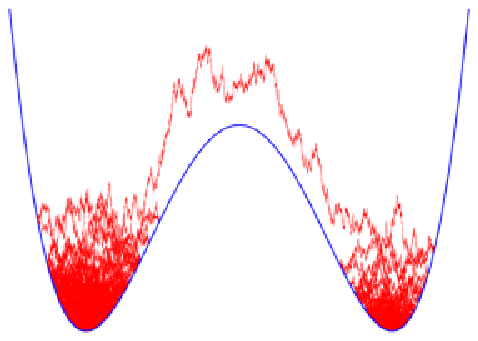}    \includegraphics{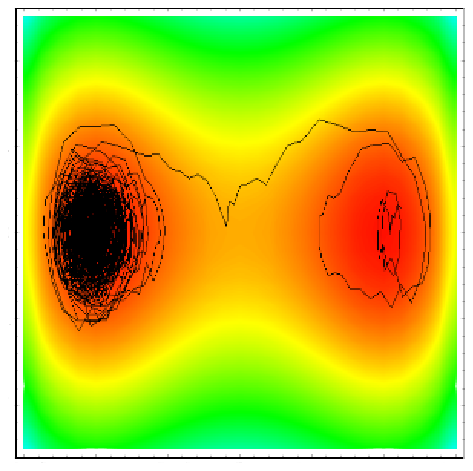} \includegraphics{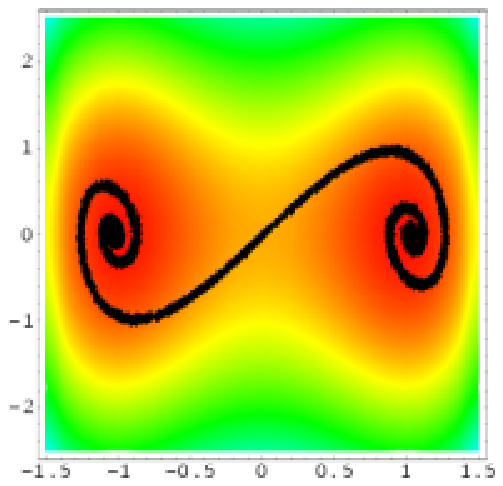}
  \caption{Probability currents for an underdamped dynamics in a 1D
    double-well potential. {\bf Left}: Typical transition between the
    two wells. {\bf Center}: Same trajectory in phase space. The
    probability current is dominated by oscillations in the
    wells. {\bf Right}: Transition current ${\bf J^{\rm t}}$ in phase
    space. The transition path is sampled uniformly. Equilibrium
    contributions are not present. }
  \label{fig:exprobcur}
\end{figure}

 For this reason, the part of the current that corresponds to
transitions between metastable states is screened by the large
contributions of the currents within metastable states (see figure
\ref{fig:exprobcur}). The probability current \eqref{eqn:currentr} and
\eqref{eqn:currentv} does not really represent the  
transition paths between metastable states. 

We can however define a {\em transition current}~\cite{TTK2006}
\begin{subequations}
\label{eqn:transcur}
  \begin{align}
    J^{\rm t}_{r_i}&= J_{r_i} +\frac{ k_B T}{m_i} \ddp{P}{v_i}\label{eqn:transcurr}\\
    J^{\rm t}_{v_i}&= J_{v_i} - \frac{ k_B T}{m_i} \ddp{P}{r_i}\;, \label{eqn:transcurv}
  \end{align}
\end{subequations}
which has two interesting properties. Firstly, this current differs
from the probability current by a divergenceless term and thus also
satisfies the continuity equation $\dot P + \nabla \cdot {\bf J^{\rm
    t}}=0$.  Fluxes out of a closed surface surrounding a metastable
state are then the same for the probability and transition
currents. The latter current thus contains the relevant information
about, for instance, transition rates. Secondly, the transition
current vanishes in equilibrium, as can be checked by
comparing~\eqref{eqn:hocur} and~\eqref{eqn:transcur}. This current
thus contains only the information relevant for the transitions
between metastable states (see figure ~\ref{fig:exprobcur}) and is
{\em not} screened by the large `equilibrium' currents within them.

Using algebra similar to that of the overdamped case, one can
show~\cite{TTK2006} (see the Appendix~\ref{app:derivation}) that the
reduced current evolves with\begin{equation}
  \label{eqn:transcurdyn}
  \frac{\partial\mathbf{J^{t}}} {\partial t}=-H_{K}\mathbf{J}^{\rm
    t}-{\mathbf{M}}\cdot \mathbf{J}^{\rm t} \;\;\; \;\;\; with \;\;\; \;\;\; 
    {\mathbf{J^t}}=\left(\begin{array}{c} J_{r_i}^t 
	   \\ J_{v_i}^t\end{array}\right)
\end{equation}
where the ${6N}\times{6N}$ matrix ${\mathbf{M}}$ is given by
\begin{equation}
  \label{m}
	{\mathbf{M}}=\left(\begin{array}{cc} 0 &
	  -\delta_{ij} \\ \frac{1}{m_i} \frac{\partial^{2}V}{\partial
	    r_{i}\partial r_{j}} &
	  \gamma\delta_{ij}\end{array}\right).
\end{equation}
Again, \eqref{eqn:transcurdyn} is a self-consistent equation for the
transition current and we shall now show how it can be simulated.

\section{Using  population dynamics to sample the transition currents}
\label{method}
The probability density $P({\bf r},{\bf v})$ is a scalar field that
can be obtained by simulating many copies of the system, evolving
their positions and velocities with the Langevin
equation~\eqref{langund}, and constructing histograms. On the
contrary, ${\bf J^{\rm t}}$ is a vector field and thus cannot be
obtained in the same way: it also requires vectorial degrees of
freedom. We first present a population dynamics that can be used to
construct the evolution of the transition current in
section~\ref{Sec:evolcurtheory} and then give the corresponding
Diffusion Monte Carlo algorithm in section~\ref{Sec:evolcuralgo}.

\subsection{Stochastic dynamics}
\label{Sec:evolcurtheory}
For systems with many degrees of freedom, the direct resolution of the
partial differential equation~\eqref{eqn:transcurdyn} yielding the
evolution of ${\bf J^{\rm t}}$ is not achievable numerically. In the
same way as the Langevin dynamics~\eqref{langund} represents an
alternative to the resolution of the Kramers equation, we can use a
stochastic dynamics that generates the current
evolution~\eqref{eqn:transcurdyn} numerically.

The explicit construction of the dynamics  was already 
presented in detail in a previous paper.~\cite{TTK2006} The idea is
to proceed in two steps. Firstly, a Langevin dynamics~\eqref{langund}
is coupled to a carefully chosen evolution (see below) of a
$6N$-dimensional vector ${\bf u}\equiv(
u_{r_1},\dots,u_{r_{3N}},u_{v_1},\cdots,u_{v_{3N}})$ in order to
account for the vectorial degrees of freedom of ${\bf J^t}$. Secondly,
the transition current ${\bf J^{\rm t}}$ is given by the following
average:
\begin{equation}
  \label{eqn:vectaverage}
	{\bf J^{\rm t}}=\int \prod_{i=1}^{6N}\; {\rm d}u_{i} \;{\bf u}\; F({\bf r}, {\bf v}, {\bf u}).
\end{equation}
where $F({\bf r},{\bf v},{\bf u})$ is a joint distribution obtained
from the Langevin dynamics, for instance by simulating a large number
of copies of the system.~\cite{foot1} 

In practice, the coupling between the vectorial and phase space
degrees of freedom is obtained by the following population
dynamics. One consider ${\cal{N}}$ copies of the system (called
`clones'), identified by positions and velocities ${\bf r}$ and ${\bf
v}$, which all carry a $6N$ dimensional unitary vector ${\bf u}$. The
dynamics of each clone is then as follows:~\cite{TTK2006}
\begin{itemize}
\item ${\bf r}$ and ${\bf v}$ evolve with the standard Langevin
  dynamics with inertia~\eqref{langund}
\item the vector ${\bf u}$ evolves with
\begin{equation}
  \label{eqn:vectdynamics}
  \dot {\bf u} = -{\bf M} \cdot {\bf u} +  {\bf u} ({\bf u}^\dagger {\bf M} {\bf u})
\end{equation}
\item each clone has a {\em birth-death rate} $\alpha=- {\bf
u}^\dagger {\bf M} {\bf u} $. This is the only way the vector ${\bf u}$ influences the dynamics.
\end{itemize}
The distribution of clones $F({\bf r}, {\bf v}, {\bf u})$ then evolves
with~\cite{G1985}
\begin{equation}
  \label{eqn:evolF}
  \frac{\partial F}{\partial t} = - H_{K} F - \sum_{i=1}^{6N}
  \ddp{}{u_i}\left[-\sum_{j} M_{ij} u_j+u_i ({\bf u}^{\dagger} {\bf M}
  {\bf u})\right] F - {\bf u}^\dagger {\bf M} {\bf u} F
\end{equation}
The first term of the r.h.s. comes from the Langevin
dynamics~\eqref{langund}, the second one from the evolution of the
vector~\eqref{eqn:vectdynamics} and the last one from the birth-death
events. The transition current is then given
by~\eqref{eqn:vectaverage}. On can indeed check that taking the time
derivative of the r.h.s of~\eqref{eqn:vectaverage} and
using~\eqref{eqn:evolF}, one recovers the evolution of the transition
current~\eqref{eqn:transcurdyn} (see previous papers~\cite{TK2004,TTK2006} for more details). 

\subsection{Algorithm}
\label{Sec:evolcuralgo}

We first present the implementation of the population dynamics of the
clones and then discuss the construction of the transition
current. The $m$-th clone is denoted by its phase-space coordinates
and vectors at time $t$,
$\mathbf{X}_{t}^{m}=\left\{\mathbf{r}^{m}(t),\mathbf{v}^{m}(t),\mathbf{u}^{m}(t)
\right\}$. We start with ${\cal{N}}_c$ clones whose positions and vectors are
 arbitrarily chosen. At every time step, the dynamics is  as
follows:
\begin{enumerate}
\item All the vectors ${\bf u}^m$ are rescaled to have a unitary norm.
\item The positions and velocities of the clones are propagated using a
  leap-frog discretized Langevin
  dynamics.~\cite{AAC2006,AA2008}
\item The vectors ${\bf u}^m$ evolve with  the
  (leap-frog discretized version of)  the following dynamics:
  \begin{equation}
    \dot u^m_i=-M_{ij}u^m_j
  \end{equation}
  Note that at after this step, the vectors are no more of unitary
  norm.
\item For each clone one records $w_m=||{\bf u}^m(t+\delta t)||$.
\item We associate to each clone $m$ a probability weight
  $\rho_m={\cal{N}}_c w_m/\sum_i w_i$ and a random number $\epsilon_m$ chosen
  uniformly in $[0,1)$. The clone is then replaced by $y_m$ copies, with
    \begin{equation} 
      y_m= \lfloor \rho_m +\epsilon_m\rfloor
    \end{equation}
    where $\lfloor x\rfloor $ is the integer part of $x$:  
    if $y_m>1$, $y_m-1$ new copies of the m-th clone are made. If
    $y_m=0$, the clone is deleted and if $y_m=1$, nothing happens. The
    population size is thus increased by $y_m-1$ if $y_m>1$ or
    decreased by $1$ if $y_m=0$.
  \item After the step 5, the population is rescaled from its current
    size ${\cal{N}}_c^e=\sum_{m=1}^{{\cal{N}}_c} y_m$ to its initial size ${\cal{N}}_c$, by
    uniformly pruning/replicating the clones.
\end{enumerate}

Steps 1 to 3 correspond to the propagation step of independent clones, whereas steps 4 to 6
correspond to selection steps. In particular, the steps 4 and 5 correctly
represents the cloning rate $\alpha=- {\bf u^\dagger M u}$ of the
previous subsection since $\frac{\rm d}{{\rm d}t} ||{\bf u}^m(t)||=-{{\bf
  u}^m}^\dagger{\bf M} {\bf u}^m$, so that $||{\bf u}^m(t+\delta t)||\simeq
\exp(-\delta t {{\bf u}^m}^\dagger {\bf M} {\bf u}^m)$.

The rescaling of the population at the step 6 can be done in many
ways. For instance, one can pick a new clone at random ${\cal{N}}_c$ times
among the ${\cal{N}}_c^e$ clones obtained at the end of the step 5. We used an
alternative approach that is less costly in terms of data
manipulations: if ${\cal{N}}_c^e>{\cal{N}}_c$, we kill ${\cal{N}}_c^e-{\cal{N}}_c$ clones chosen
uniformly at random among the ${\cal{N}}_c^e$ obtained at the end of the step
5. Conversely, if ${\cal{N}}_c^e<{\cal{N}}_c$, we choose uniformly at random
${\cal{N}}_c-{\cal{N}}_c^e$ clones and duplicate them. Note that even though the
clones evolve independently during the propagation step, their
dynamics are correlated because the deleted clones are replaced by the
duplicated ones at the selection steps. When the probability weights
of the clones are all equal, we have $\rho_m=1$ and $y_m=1$. As a
result, the population is left unchanged. Conversely, when the clone
weights take distinct values, clones with small weights are likely to
be replaced by those with large weights.

There are many ways of implementing the resampling of the population
(steps 5-6), well documented in the literature on Diffusion Monte
Carlo.~\cite{MJL2007,C2007} In particular, it
could be advantageous to do the resampling only every $n$ time steps,
where $n$ is tuned to ensure ergodicity in phase space, i.e. to
achieve enhanced sampling towards the unstable regions where saddles
are located.

Since the clones move in phase space with a Langevin dynamics, it can
be surprising that they converge rapidly to the reaction paths,
i.e. that they explore efficiently the transition states. This can
however be understood by noting that their dynamics (without taking
the averages \eqref{eqn:vectaverage}) is the so-called \emph{Lyapunov
Weighted Dynamics}~\cite{TK2006} which is used to bias the Langevin
dynamics in favor of chaotic trajectories. The clones will then tend
to `reproduce' favorably in the neighborhood of saddles, which are
particularly chaotic regions of phase space, and to die in wells. This
generates an `evolutionary pressure' that helps the clone escape from
metastable states and find the reaction paths. 

As mentioned before, this dynamics does not provide {\em directly} the
transition current and one still has to construct the averages
\eqref{eqn:vectaverage}. This can be difficult and clever methods to
do so were discussed in the literature, for instance by Mossa and
Clementi who studied the folding of chain of
aminoacids.~\cite{MC2007} The difficulty is connected to the
well-known sign problem: large population of clones with arrows
pointing in opposite directions cancel in the average but can
numerically screen smaller asymmetric distribution that contains the
information relevant for the transition current. One can however show
that if one starts from a population of clones uniformly spread over a
reaction path separating two metastable states and pointing in the
same direction, the time taken for the sign problem to occur is of the
order of the tunneling time through the barrier (see
Appendix~\ref{app:ex1d}). In the following we will always simulate
  much shorter times and omit the averaging steps to simply look at
  the distribution of clones. This distribution often suffices to
  locate the reaction paths, as can be seen on figure
  \ref{fig:exprobcur} for the double well potential. To extract
  further information, for instance regarding the reaction rates, we
  would need to do the averages, as was done in Ref.~\onlinecite{MC2007}, but
  this is beyond the scope of this article.

\section{Applications}
\label{sec:LJ38}

\subsection{LJ$_{38}$ cluster}

\label{LJ38Intro}
We now turn to the study of transitions between metastable states in
the 38-atom Lennard-Jones cluster, a benchmark model system that has
been extensively investigated in the
past.~\cite{DMW1999b,DMW1999,NCFD2000,CNFD2000,AAC2006}  This system
has a complex potential energy landscape organized around two main
basins: a deep and narrow funnel contains the global energy minimum, a
face-centered-cubic truncated octahedron configuration (FCC), while a
separate, wider, funnel leads to a large number of incomplete Mackay
icosahedral structures (ICO) of slightly higher energies.

Although the lowest potential energy minimum corresponds to the FCC
structure, the greater configurational entropy associated to the large
number of local minima in the icosahedral funnel make this second
configuration much more stable at higher temperatures. As temperature
increases, this system thus undergoes the finite-size counterpart of
several phase transitions. First, a solid-solid transition occurs at
$T_{ss}=0.12 \frac{\epsilon}{k_{B}}$ when the octahedral FCC structure
gives place to the icosahedral ones. At a slightly higher temperature,
$T_{sl}=0.18 \frac{\epsilon}{k_{B}}$, the outer layer of the cluster
melts, while the core remains of icosahedral
structure.~\cite{MF2006} This `liquid-like' structure, also referred
to as anti-Mackay in the literature, then completely melts around
$T_{sl}=0.35 \frac{\epsilon}{k_{B}}$.~\cite{MF2006}

The numerical study of this system is challenging: global optimization
algorithms have failed to find its global energy minimum for a long
time~\cite{W2003} and direct Monte Carlo sampling fails to
equilibrate the two funnels. The study of the equilibrium
thermodynamics of this system required more elaborate algorithms such
as parallel tempering,~\cite{NCFD2000,CNFD2000,MF2006} basin-sampling
techniques,~\cite{BWC2006} Wang-Landau approaches~\cite{PCABD2006} or
path-sampling methods.~\cite{AAC2006,MP2007,AM2010}

More recently, the dynamical transitions between the two basins has
been studied following various approaches. The interconversion rates
between the FCC and ICO structures have been computed using Discrete
Path Sampling.~\cite{W2002,W2004,BW2006} This elaborate algorithm
relies on the localization of minima and saddles of the potential
energy landscape, using eigenvector following, and then on graph
transformation~\cite{TW2006} to compute the overall transition rate
between two regions of phase space.  To the best of our knowledge,
this is the most successful approach as far as computing reaction
rates in LJ$_{38}$ is concerned.~\cite{TW2006} However, the 
numerical methods involved are quite elaborate, require considerable expertise
and have a number of drawbacks, all deriving from the fact that it is
based on the harmonic superposition approximation and the theory of
thermally activated processes. It thus requires any intermediate
minima between the two basins to be equilibrated and this is only
possible for small enough systems at low temperatures.~\cite{W2002}
More importantly, when the difficulty in going from one basin to the other is
due to entropic problems, as is the case for instance in hourglass
shaped billiards, then the knowledge of minima and saddles of the
potential energy landscape is  not sufficient.

Another attempt to study the transitions between the two funnels of
LJ$_{38}$ relies on the use of transition path sampling.~\cite{MP2007}
Because of the number of metastable states separating the two main
basins, the traditional shooting and shifting algorithm failed here,
despite previous success for smaller LJ clusters.~\cite{DBC1998} The
authors thus developed a two-ended approach which manages to
successfully locate reaction paths between the two basins: they
started from a straight trial trajectory linking the two minima, and
obtained convergence towards trajectories of energies similar to those
obtained in the Discrete Path Sampling approach.~\cite{MP2007}
Although the authors point out the lack of ergodicity in the sampling
within their approach and the sensitivity on the `discretization' of
the trajectories, this is nevertheless a progress and the main drawback
remains the high computational cost (the work needed $10^5$ hours of
cpu time) to obtain such converged trajectories. In contrast, the
simulations we present below required less than $10^2$ hours of cpu
time.

\subsection{The LJ$_{38}$ cluster and bond-orientational order parameters}
Before presenting our simulations results, we give some technical
details on the LJ$_{38}$ system and on the visualization of the different
metastable states. The Lennard-Jones potential is given by the
expression
\begin{equation}
  V\left(\left\{ \mathbf{r}^{j}\right\}
  _{j=1,...,N}\right)=4\epsilon\sum_{j<k}
  \left[\left(\frac{\sigma}{r_{jk}}\right)^{12}-\left(\frac{\sigma}{r_{jk}}\right)^{6}\right]
\end{equation}
where $\mathbf{r}^{j}=(r^j_{x},r^j_{y},r^j_{z})$ is the position of
the $j$-th atom, $r_{jk}=\left|\mathbf{r}^{j}-\mathbf{r}^{k}\right|$
is the distance between atoms $j$ and $k$, $\epsilon$ is the pair well
depth and $2^{1/6} \sigma$ is the equilibrium pair separation. In
addition, all the particles are confined by a trapping potential that
prevents evaporation of the clusters at finite temperature
(i.e. particles going to infinity). If the distance between the
position ${\bf r}$ of a particle and the center of the trap ${\bf
  r_c}$ exceeds $2.25 \sigma$, then the particle feels a potential
$|{\bf r}-{\bf r_c}|^3$.  LJ reduced units of length, energy and mass
($\sigma=1, \epsilon=1, m=1$) will be used in the sequel so that the
time unit $t=\sigma\sqrt{m/\epsilon}$ is set to $1$.

\begin{figure}
 \includegraphics{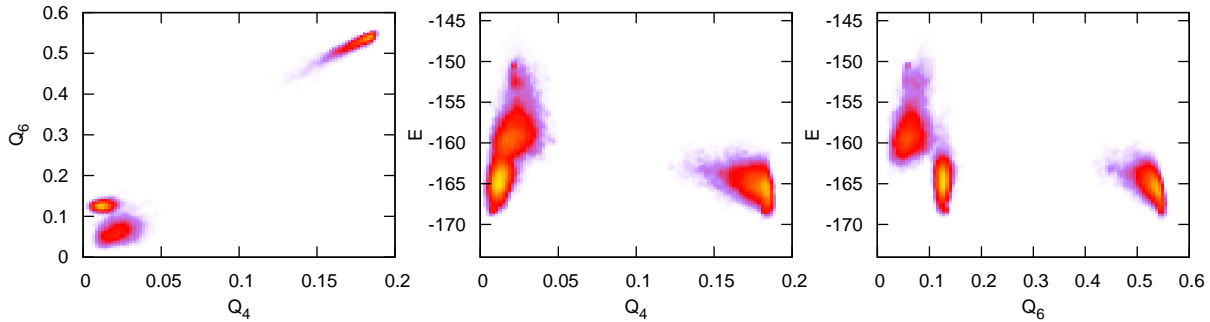}
  \caption{Short MD simulations were run to give an impression of the
    spread in the (Q6,Q4,E) space of each `phases'. The simulation
    time was short enough that no tunneling between the phases was
    observed. The temperature was set to $T=0.15$. The positions of
    the phases barely move in the $(Q_4,Q_6)$ plane when the
    temperature changes, although their spreading does. The kinetic
    energy however shifts when the temperature changes, and is roughly
    proportional to $N kT$ where $N$ is the number of degrees of
    freedom. }
  \label{fig:bassinsMD}
\end{figure}

Rather than listing the 228 degrees of freedom of the atomic cluster,
configurations are traditionally described using the $Q_{l}$
bond-orientational order parameters
\cite{SNR1981,SNR1983} that allow to differentiate
between various crystalline orders
\begin{equation}
Q_{l}=\left(\frac{4\pi}{2l+1}\frac{1}{{{N}}_{b}^{2}}\sum_{m=-l}^{l}\left|Y_{lm}\left(\theta_{jk},\phi_{jk}\right)\right|^{2}\right)^{1/2}\;,
\end{equation}
where the $Y_{lm}(\theta,\phi)$ are spherical harmonics and
$\theta_{jk}, \phi_{jk}$ are the polar and azimuthal angles of a
vector pointing from the cluster center of mass to the center of the
\textit{(j,k)} bond which connects one of the $N_{b}$ pairs of
atoms.~\cite{foot2} The parameter $Q_{4}$ is
often used to distinguish between the icosahedral and cubic
structures, for which it has values around 0.02 and 0.18
respectively.~\cite{NCFD2000} $Q_4$ however does not distinguish between the
icosahedral and the liquid-like phase and one thus often uses $Q_6$, for which
FCC, icosahedral  and the liquid-like phase take values around $0.5$, $0.13$
and $0.05$,~\cite{DMW1999} respectively. To show the spread of the
various basins in the $(Q_4,Q_6)$ plane, we ran several molecular dynamic
simulations, long enough to equilibrate within each basin but short
enough so that one does not see tunneling (see
figure~\ref{fig:bassinsMD}).

Although the whole temperature scale is interesting, the challenging
part from a computational point of view is the low-temperature regime
where ergodicity is difficult to achieve. Below, we show the results
of our algorithm for three temperatures: $T=0.12 \epsilon/k_B$,
$T=0.15 \epsilon/k_B$ and $T=0.19 \epsilon/k_B$ that span the ranges
around the solid-solid and partial melting transitions.

\subsection{Simulations}

Given the high dimensionality of the system, it is difficult to follow
 the evolution of all the coordinates of the clones  in order to know if they have
localized interesting structures. Instead, we proceed as follows: we
plot the evolution of the average over the clone population of $Q_4$,
$Q_6$ and $E$ as a function of the simulation time and we frequently
store the positions and velocities of all the clones.

 If we see a plateau in $Q_4(t)$, $Q_6(t)$ and $E(t)$, two cases are
 possible: either the clones have converged to a reaction path, or
 they are stuck in a metastable basin. In order to distinguish the two
 situations, we run an auxiliary short molecular dynamic simulation
 ({\em without cloning}) starting from the positions and velocities of
 the clones. The duration of this auxiliary simulation is long enough to
 observe relaxation into the metastable basins, but much shorter than
 the transition times. If the clones evolve away from the region they
 had populated in phase-space, we know they had found a reaction path
 and the auxiliary MD simulation converges to the metastable basins
 connected by this reaction path. If on the other hand the clones do
 not evolve away, we know that they had been stuck in some local
 basin. In such a case we can change two parameters to enhance the
 sampling of the phase space: the number of clones and the friction
 $\gamma$ (see below for more details). The time step is always
 $\delta t = 0.01$. Note that this procedure could be automated, but
 the way to do so is let for future work.

In principle, any observable that can measure whether the population
of clones splits in two separate sub-populations after a short
Langevin dynamics would be suitable.  If the clone population splits
in two subpopulations with the same $Q_4$, $Q_6$ and $E$, we may fail
to detect the corresponding barrier. However, this coincidence would
be extremely unlikely.

Last, in addition to help us localize barriers, these short Langevin
simulations allows us to explore the true dynamics close to a
particular transition states.

\subsubsection{$T=0.15$}
We first ran several simulations at $T=0.15$, where the most stable
state is the MacKay icosahedral minimum (ICOm) while the liquid-like
phase (ICOam) and the FCC basin are metastable.

Starting from the ICOm basin with ${\cal{N}}=200$ clones and a low friction
$\gamma \delta t=10^{-3}$, the clones rapidly find ($t\sim 1500$) a transition
path to the liquid-like phase ICOam. Later on ($t\sim 3500$) an
activated event bring the clones to another reaction path that points
towards the FCC funnel. These times have to be compared with the
transition time between the ICOm and FCC basins that was previously
evaluated in the literature at roughly $10^7$.~\cite{W2002} Note that
each barrier or path act as a metastable state for the cloned
dynamics and it is by activation that the population jumps from one
barrier to another. Running the same dynamics with a larger number of clones (${\cal{N}}=600$)
tends to stabilize the first barrier so that one has to wait longer to
see the transition to the second one.

Starting from the FCC minimum with the same number of clones and at
the same friction results in the clones rapidly going out of the FCC
funnel and falling in the amorphous zone at the entrance of the
icosahedral funnel.~\cite{NCFD2000} A reaction path is {\em followed}
by the clones but not {\em maintained}. To stabilize this reaction
path, we increased the number of clones and the friction. The effect
of the former is mostly to slow down the dynamics while the latter
allow the clones to populate the reaction path more uniformly. For
${\cal{N}}=600$ clones and $\gamma \delta t =1$, the population of clones indeed
stabilizes the reaction path leading from the FCC basins to the
entrance of the icosahedral funnel. The reason why we need more clones
to stabilize this barrier than the ones in the icosahedral funnel is
probably that the former is more flat and spread than the latter
ones~\cite{DMW1999} and thus requires a larger number of clones to be
sampled uniformly.

\begin{figure}
\includegraphics{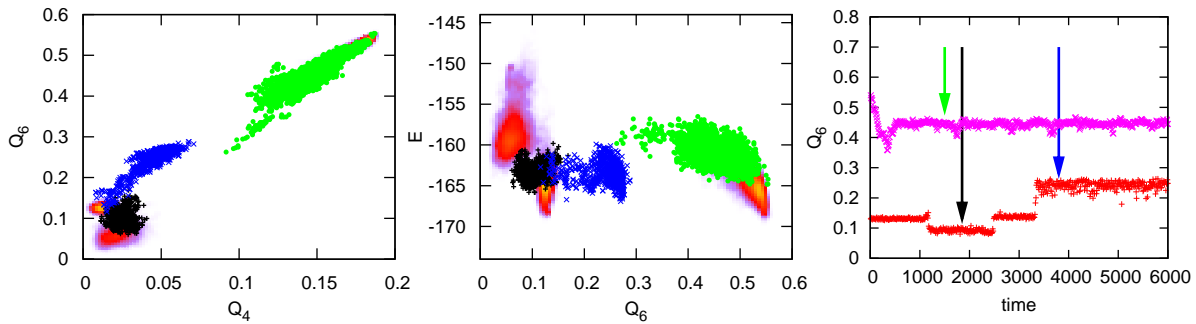}
  \caption{{\bf Left and Center:} Positions of the clones starting
    from the ICOm and FCC minima in the $(Q_4,Q_6)$ and $(Q_6,E)$
    plans at $T=0.15$. The clones starting from the icosahedral basin first find
    the barrier between ICOm and ICOam (black symbols, $t=1850$).
    They then fall back in the ICO basin before finding a path that
    points towards the FCC funnel (blue symbols, $t=3500$). Starting
    from FCC, the $600$ clones find a path that leads toward the
    icosahedral funnel (green symbols, $t=1500$). {\bf Right:} We plot
    $Q_6$ as a function of time for the clones starting from the ICOm
    basins (red symbols) and the FCC basin (magenta symbols). Arrows
    indicates the time at which the snapshots shown in the left and
    center panel are taken.}
  \label{fig:cloneT015}
\end{figure}

All these results are plotted on figure \ref{fig:cloneT015}, in which
we show the three basins ICOm, FCC and ICOam obtained from the initial
MD simulations (see figure~\ref{fig:bassinsMD}) and the positions of
the clones in the $(Q_4,Q_6)$ and $(Q_6,E)$ plans at different
simulation times.

\begin{figure}
\includegraphics{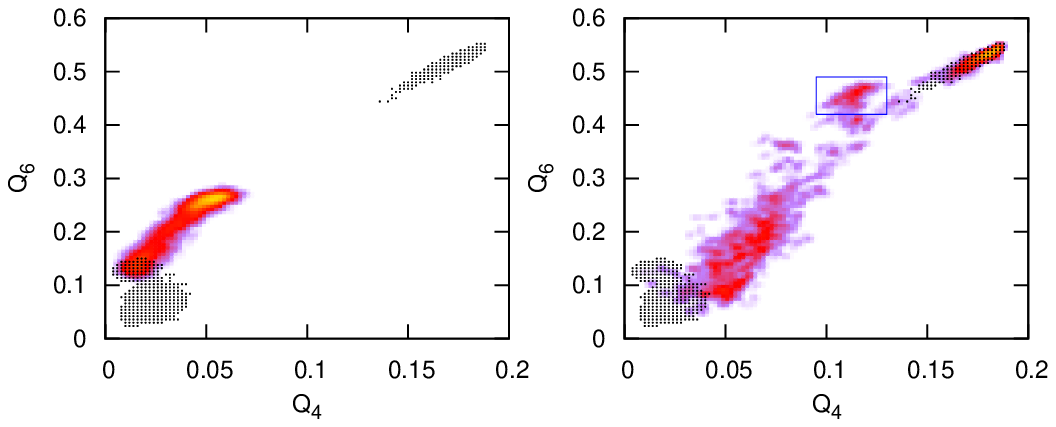} \hspace{.5cm}\includegraphics[totalheight=4.5cm]{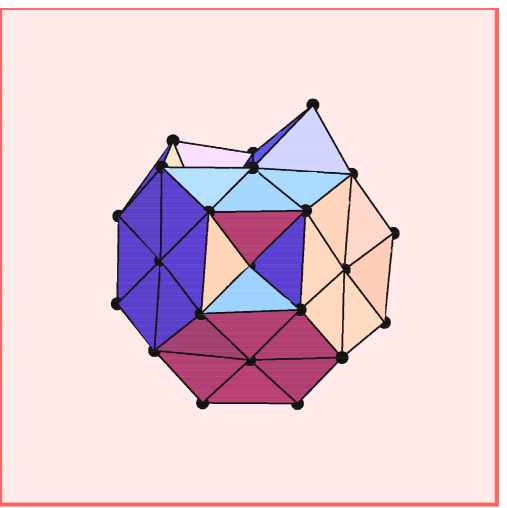}
  \caption{Histograms made at the end of short MD simulations at
    $T=0.15$ started from the clones positions at the times indicated
    by the green and blue arrows in figure~\ref{fig:cloneT015}. The
    gray-dotted regions correspond to the equilibrium MD simulations
    of the three basins ICOm, ICOam and FCC. {\bf Left}: MD
    simulations started from the stationary structures found by the
    clones in the ICO funnel fall either back into the ICO basin or in
    a metastable basin around $(Q_4,Q_6)=(0.3,0.05)$ that corresponds
    to an amorphous structure at the entrance of the ICO funnel. {\bf
      Center}: The clones starting from the stable structure found in
    the FCC funnel fall either back in the FCC basin, or in a faulty
    FCC metastable state (blue rectangle) or in the ICO funnel,
    which. Both structures thus correspond to reaction paths. {\bf
      Right}: Position of the surface atoms of a clone that has fallen
    in the faulty FCC configuration after the short MD. This
      figure was made using a Mathematica Spreadsheet that can be
      downloaded at  \url{http://www-wales.ch.cam.ac.uk/~wales/make_frames.nb}.}
  \label{fig:T015HistoClones}
\end{figure}

To identify the various metastable basins connected by the clones, we
ran several short MD simulations starting from the two long-lived
plateaux (blue and green arrows in the right panel of
figure~\ref{fig:cloneT015}). Histograms made from these MD runs are
shown in figure~\ref{fig:T015HistoClones}. They show that the clones
going out of the ICO basin find barriers toward the amorphous region
at the entrance of the ICO funnel while the ones starting from the FCC
minimum find a reaction path between the FCC basin and the icosahedral
funnel. Interestingly, this path  goes through a faulty FCC basin located
around $(Q_4,Q_6)=(0.12,0.45)$ that has been  previously reported in the
literature.~\cite{NCFD2000,AAC2006}

The clones have thus found reaction paths pointing out of their
starting funnels. The clones starting from the FCC basins find a
reaction path that leads into the icosahedral funnel while the one
started from the ICOm basin still remain in the icosahedral
funnel. This could be explained by the fact that at this temperature
ICOm is the {\em stable} state while FCC is only metastable so that
the barrier ICO$\to$FCC has to be harder to access than the one from
the FCC side. Running short MD starting from the clones positions
reveal intermediate metastable basins, either a faulty FCC or
amorphous structures.

\subsubsection{$T=0.12$}

This is the coexistence temperature between the ICOm and FCC
minima. At such a low temperature, more and more secondary barriers
play a role so that the transition between ICO and FCC becomes more
and more complex. From the point of view of the time evolution of the
transition current, this means that there are more and more metastable
states for the clone dynamics.

Starting from the ICOm basin with 600 clones and $\gamma \delta t=10^{-3}$, we
once again locate the barrier between ICOm and liquid-like phase
ICOam. At this temperature this barrier is long-lived and we do not
locate the one previously found at $T=0.15$ that points toward the
entrance of the icosahedral funnel.

Starting from the FCC basin with $\gamma \delta t =0.02$, the $600$ clones find
several barriers that constitute a multi-step reaction path toward the
icosahedral funnel. Once again, the larger the friction the longer the
clones spend on intermediate barriers.\cite{foot3} 

\begin{figure}
\includegraphics{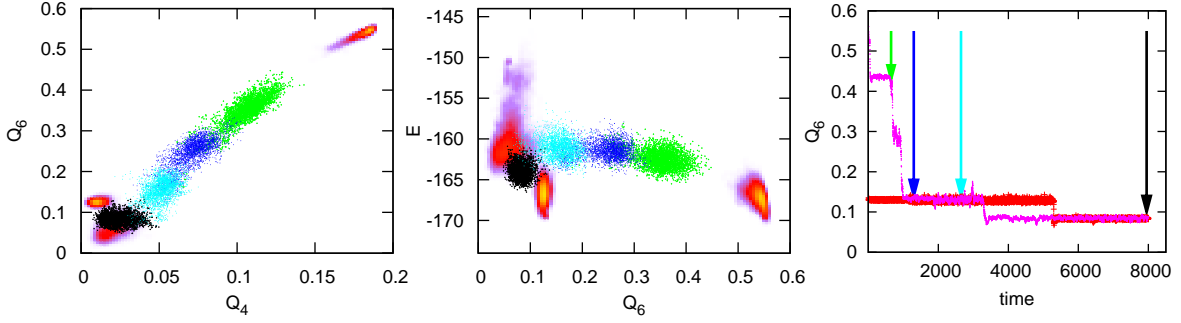}
  \caption{{\bf Left and Center}: Positions of the clones starting
    from the ICOm and FCC minima in the $(Q_4,Q_6)$ and $(Q_6,E)$
    plans at $T=0.12$. The clones starting from ICO find the barrier
    between ICOm and ICOam (black symbols, $t=8000$). Starting from
    FCC, the clones find a succession of barriers that leads toward
    the icosahedral funnel (green symbols at $t=650$, blue symbols at
    $t=1300$ and cyan symbols at $t=2650$). {\bf Right:} We plot $Q_6$
    as a function of time for the clones starting from the ICOm basins
    (red symbols) and the FCC basin (magenta symbols).}
  \label{fig:clonesT012}
\end{figure}

The position of the clones corresponding to the successive metastable
barriers are shown in figure~\ref{fig:clonesT012}. Note that the
typical times needed for locating the barriers are of the order of
$10^3$, that is {\em seven orders of magnitude smaller} than the reaction times between
ICOm and FCC, which is of the order of $10^{10}$.~\cite{W2002}

\begin{figure}
\includegraphics{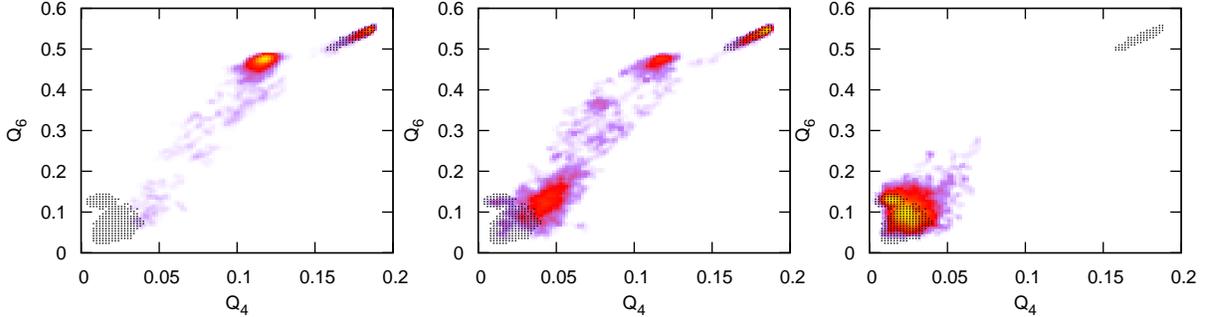}
  \caption{The color codes correspond to MD simulations at $T=0.12$
    started on the green (left), blue (center) and black (right)
    arrows in figure~\ref{fig:clonesT012}. {\bf Left:} Starting from
    the first stationary structure found in the FCC funnel, the clones
    relaxes mostly in the FCC basin and in the faulty FCC
    configuration shown in~\ref{fig:T015HistoClones}. {\bf Center} The
    second barrier is close to the commitor between the ICO and the
    FCC funnel: the clone population relaxes almost equally in both
    funnel ($57\%$ of the clones fall back in the icosahedral funnel
    while $43\%$ enter the fcc funnel). {\bf Right} Clones started
    from the barrier between ICOm and ICOam populate both basins. Note
    that the relaxation is much slower than for the other barrier
    because of the entropic nature of this barrier. }
  \label{fig:T012MDafterClones}
\end{figure}

Running short MD simulations starting from the clone positions on the
barriers and constructing the corresponding histograms reveals various
intermediate metastable basins in the ICO and FCC funnels (See
figure~\ref{fig:T012MDafterClones} ). The fact that $Q_4$ and $Q_6$
are not good reaction coordinates is confirmed in this figure: the
first plateau (green points on figure~\ref{fig:clonesT012}) seems to
be {\em after} the faulty configuration when going from the FCC funnel
to the icosahedral one but the MD starting from this barrier falls
into the faulty configuration and the FCC basin, which seems to
indicate that this barrier is a reaction state between the FCC and the
faulty configuration. There is then a second barrier between the
faulty FCC and the ICO funnel (blue dots in
figure~\ref{fig:clonesT012}). A last barrier leads to the amorphous
region that separates the liquid-like phase and ICOm minimum. Note
that in these regions the MD simulations are not very helpful because
the transition between ICOm and ICOam has an entropic nature so that
it is difficult to relax into the basins. The clones, however,
successfully identify the barrier between these two states.

\subsubsection{$T=0.19$}

This temperature is very close to the transition between ICO and
liquid-like phase. As shown by free-energy studies, the barrier between
the FCC and the ICO funnels is very low and the FCC basin is rather
unstable.~\cite{DMW1999} Clones starting from the FCC basin do not
stabilize on any structure because there is no proper `rare barrier' and
MD simulations starting from FCC immediately falls into the
icosahedral funnel.~\cite{DMW1999}

Starting $100$ clones from the ICO basin at $\gamma \delta t = 0.01$, they rapidly
find a barrier connecting to the liquid-like phase. Later on,
activated events lead the clones to locate a reaction path leading
towards the FCC funnel. Starting MD from this barrier show that the
clones relax into the FCC and ICO funnel.

\begin{figure}
\includegraphics{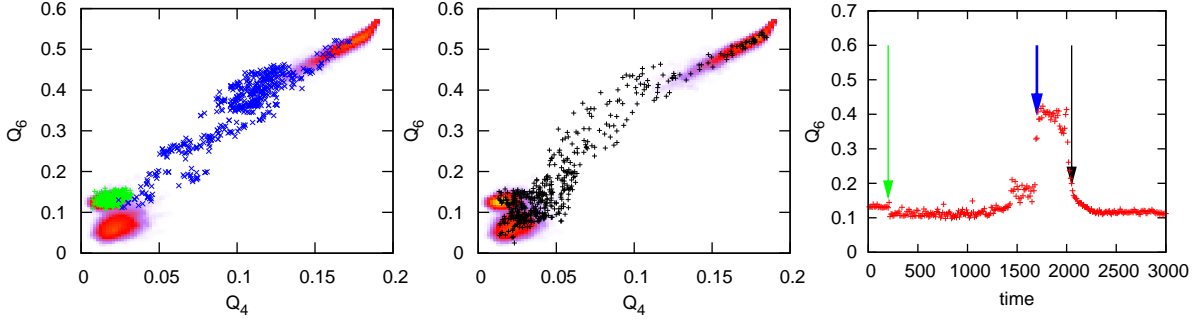}
  \caption{{\bf Left}: $100$ Clones are started at $T=0.19$ in the ICOm
    basin where they spend some time (first first time steps after
    $t=200$, green dots) before locating the barrier toward the FCC
    funnel (first five time steps after $t=1700$, blue crosses). {\bf
      Center:} When the clones have found the barrier ($t=2000$) a
    standard MD starts and relaxes as it should into the two funnels
    (black dots, five first snapshots after $t=2050$). {\bf Right:}
    Evolution of $Q_6(t)$. The cloning is stopped at $t=2000$ and a
    normal MD follows.}
  \label{fig:clonesT019}
\end{figure}

As mentioned above, it is hard for the clones to stabilize because the FCC
funnel is barely metastable and the barrier crossed while going from
FCC to ICO is rather flat at this temperature.
 It is thus quite surprising that they nevertheless manage
to do so while starting from the ICO basin. If one starts from the FCC
funnel, the clones almost immediately fall into the ICO funnel and
then from there can locate the barrier again,  but we were not able to
stabilize the barrier when coming from the FCC basin. This might be
due to the fact  that  clones stabilize reactions that take
place on long time-scale (ICO$\to$FCC), but not short-time relaxations
(FCC$\to$ICO).

\subsection{$T=0.05$: annealing the cloned system}

\begin{figure}[h!]
  \begin{center}\includegraphics{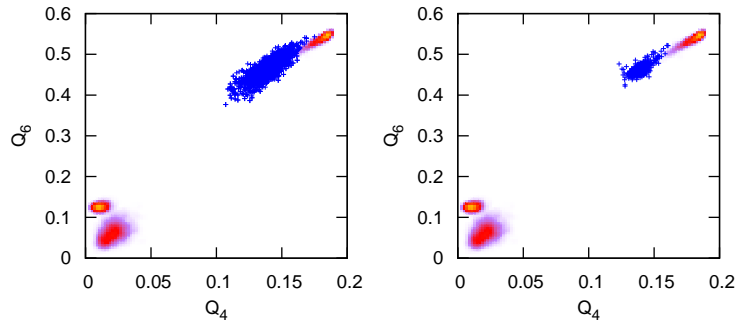}\end{center}
  \caption{{\bf Left}: Result of a simulation run at T=0.12 with 600
    clones, starting from the FCC minimum, with $\gamma \delta t =0.6$ {\bf
      Right:} Starting from the end point of the simulations at
    T=0.12, we run a standard cloning simulation at $T=0.05$. After a
    time $t=1790$ the 600 clones are still on the structure that had
    localized at $T=0.12$, which is thus very stable.}
  \label{fig:annealing}
\end{figure}

If one starts at such a low temperature from one of the various
metastable basins, the clones remain trapped for a time longer than
the simulation time. One can however use a temperature annealing to
locate the barriers. If one starts the cloning simulation at $T=0.12$
or higher, it is quite easy, as we saw above, to localize the
barriers. The temperature can then be decreased to $T=0.05$ and the
clones remains on the structure that were localized at a higher
temperature (see figure~\ref{fig:annealing}).

\section{Conclusions}

The algorithm we have discussed in this paper may be characterized as
one that simulates the evolution in time of the current distribution,
rather than that of the configuration.  Because the time for the
escape current to be established is often much smaller than the
passage time itself, the method is able to find the transition paths
very efficiently.

The method has several attractive features:

{\em i}) It does not require any previous knowledge of the relevant
reaction coordinates. On the other hand, if an approximation of the
reaction path is known {\em a priori}, one may always start the clones
along this path, and they will populate the true current distribution
in a shorter time.

{\em ii}) Because the target of the dynamics is the reaction path
distribution itself, one may perform simulated annealing in path
space: first populating the reaction path corresponding to relatively
high temperature, and then refining it to the lower, target
temperature. Repeated annealing can also be used to locate several
competing barriers in system with multiple reaction mechanisms.
 
 {\em iii}) The reaction current vanishes between mutually thermalized
 regions.~\cite{foot4} This is why
 at longer times, the system converges to the barriers that take
 longer to cross, {\em irrespective of whether they are of entropic or
   energetic nature}. This may be an advantage in cases in which the
 energy landscape is not in itself dominant, but rather the
 multiplicity of paths dominates.
 
{\em iv}) { The construction of the transition current
  and the cloning algorithm also applies for non-equilibrium systems
  where the forces derive locally, but not globally from a potential, such as a system 
  with leads at the edges having  a  potential difference.
 Reaction paths between non-equilibrium
  metastable states, which cannot be described in term of a free
  energy, may be studied in the same way. The only difference is that the
  average~\eqref{eqn:vectaverage} does not vanish in the long time
  limit and converges instead to the steady-state transition current.}

Note that since the method does not require the knowledge of the
reaction coordinate, it could be used efficiently in systems with
competing reactions where one does not know {\em a priori} the end
points of the reaction paths. This would for instance be particularly
interesting when studying the crystallization of suspensions of
oppositely charged colloids.~\cite{SVFD2007,P2009}

In principle, the reaction time may be expressed directly in terms of
the (unnormalized) reaction current.  It may also be recovered from
the weights carried by the clones, which may possibly be achieved from
importance sampling in a Lyapunov-weighted ensemble of
trajectories.~\cite{GD2010} However, the method, as it stands does not
allow one to calculate the reaction {\em time} with great precision,
due to the exponential nature of the timescale. Further work is
required in this direction.

Acknowledgement: We thank the ESF programme ``SimBioMa - Molecular
Simulations in Biosystems and Material Science''(M.P.)  and EPSRC
Ep/H027254 (J.T.) for funding, D. Wales and C. Valeriani for repeated
fruitful discussions and F. Calvo and D. Wales for giving us several numerical
routines. 

\section{Appendix A}\label{app:derivation}

We report here the derivation of the time evolution equation
\eqref{eqn:transcurdyn} for the transition current in the underdamped
(i.e., Kramers) case of section \ref{sec:kram}.

The classical Kramers probability current $\mathbf{J}$, presented in equations \eqref{eqn:transcur} , can indeed be written, as in the Fokker-Planck case (Eq. \eqref{eqn:newcur}), in the operatorial form $\mathbf{J}(\mathbf{r},\mathbf{v},t)=\hat{\mathbf{J}}P(\mathbf{r},\mathbf{v},t)$, where the components of the current Kramers operator 
 are
\begin{subequations}
\begin{align}
\label{a3}
\hat{J}_{r_{i}}= & {v_{i}}\\
\hat{J}_{v_{i}}= & -\left(\beta^{-1}\frac{\gamma}{m_{i}}\frac{\partial}{\partial v_{i}}+\gamma v_{i}+\frac{1}{m_{i}}\frac{\partial V}{\partial r_{i}}\right)\label{a3b}\end{align}\end{subequations}
In equations \eqref{eqn:transcur} of section \ref{sec:kram}, a \textit{transition current} $\mathbf{J}^{t}$ has been introduced, which can be expressed in turn with operators as 
\begin{equation}
\mathbf{J}^{t}=\mathbf{\hat{\mathbf{J}}}^{t}P=\left(\hat{\mathbf{J}}+\hat{\mathbf{T}}\right)P
\end{equation}
where $\mathbf{\hat{\mathbf{J}}}$ is the Kramers current operator reported above, giving the usual
{\em phase-space}  current $\mathbf{J}$, and the `transition' operator 
\begin{eqnarray}
\label{a4}
\hat{\mathbf{T}}_{r_i}&=&\frac{1}{\beta m_i}\frac{\partial}{\partial v_{i}}\nonumber \\
\hat{\mathbf{T}}_{v_i}&=&-\frac{\partial}{\partial r_{i}}
\end{eqnarray}
$\hat{\mathbf{T}}P$  is a divergenceless term. 
 As already remarked in section \ref{sec:kram}, the transition current still satisfies the continuity equation \begin{equation}
\frac{\partial P}{\partial t}=-\nabla_{\mathbf{r},\mathbf{v}}\cdot\mathbf{J}^{t}
\end{equation}
thanks to the divergenceless of  $\hat{\mathbf{T}}P$. We have introduced here the phase-space divergence
$\nabla_{\mathbf{r},\mathbf{v}} \equiv (\nabla_{\mathbf{r}},\nabla_{\mathbf{v}})$

As in section \ref{sec:fp}, we proceed now in deriving explicitly the time evolution equation of $\mathbf{J}^{t}$.  Multiplying both sides of the continuity equation above (indeed identical to the Kramers equation \eqref{2.14}) by the transition current operator leads to 
\begin{equation}
\label{a1}
\hat{\mathbf{J^t}} \frac{\partial P}{\partial t}=-\hat{\mathbf{J^t}}\nabla_{\mathbf{r},\mathbf{v}}\cdot\hat{\mathbf{J^t}}P
\end{equation}
On the l.h.s. the transition current operator can be commuted with the  time derivative. The r.h.s. of \eqref{a1} can be rewritten with commutators as
\begin{equation}
\label{a2}
\hat{\mathbf{J}}^{t}\nabla_{\mathbf{r},\mathbf{v}}\cdot\hat{\mathbf{J}}^{t}=\nabla_{\mathbf{r},\mathbf{v}}\cdot\hat{\mathbf{J}}\left(\hat{\mathbf{J}}+\hat{\mathbf{T}}\right)+\left[\hat{\mathbf{J}},\nabla_{\mathbf{r},\mathbf{v}}\cdot\hat{\mathbf{J}}\right]+\left[\hat{\mathbf{T}},\nabla_{\mathbf{r},\mathbf{v}}\cdot\hat{\mathbf{J}}\right]
\end{equation}
using ~\eqref{a1} and the zero divergence property of $\hat{\mathbf{T}}$. 

Resorting to definitions of the current operator $\hat{\mathbf{J}}$ and the transition operator $\hat{\mathbf{T}}$ given in ~\eqref{a3},~\eqref{a3b} and ~\eqref{a4}, explicit expressions for commutators in ~\eqref{a2} can be recovered with straightforward algebra: the term $\left[\hat{\mathbf{J}},\nabla_{\mathbf{r},\mathbf{v}}\cdot\hat{\mathbf{J}}\right]$ gives
\begin{subequations}\label{a5}
\begin{align}
\left[\hat{J}_{r_{a}},\nabla_{\mathbf{r},\mathbf{v}}\cdot\hat{\mathbf{J}}\right]P &=-\sum_{i}\delta_{ia}\left(J_{v_{i}}-\frac{\gamma}{\beta m_i}\partial_{v_{i}}\right)P\\
\left[\hat{J}_{v_{a}},\nabla_{\mathbf{r},\mathbf{v}}\cdot\hat{\mathbf{J}}\right]P &=\sum_{i}\delta_{ia}\left(\gamma J_{v_{i}}-\frac{\gamma}{\beta m_i}\partial_{r_{i}}\right)P-\sum_{i}\frac{1}{m_i}\frac{\partial^{2}V}{\partial r_{i}\partial r_{a}}v_{i}P\end{align}
\end{subequations} while $\left[\hat{\mathbf{T}},\nabla_{\mathbf{r},\mathbf{v}}\cdot\hat{\mathbf{J}}\right]$ can be expressed as 
\begin{subequations}\label{a7}
\begin{align}
\left[\hat{T}_{r_{a}},\nabla_{\mathbf{r},\mathbf{v}}\cdot\hat{\mathbf{J}}\right]P&=\sum_{i}(\beta m_i)^{-1}\delta_{ia}\left(\partial_{r_{i}}+\gamma\partial_{v_{i}}\right)P\\
\left[\hat{T}_{v_{a}},\nabla_{\mathbf{r},\mathbf{v}}\cdot\hat{\mathbf{J}}\right]P&=\sum_{i}(\beta {m_i}^2)^{-1}\frac{\partial^{2}V}{\partial r_{i}\partial r_{a}}\partial_{v_{i}}P\end{align}\end{subequations} 
Inserting now ~\eqref{a5} and ~\eqref{a7} in the r.h.s. of Eq.~\eqref{a1} yields
\begin{subequations}\label{a8}
\begin{align}
(\hat{J}_{r_{a}}+\hat{T}_{r_{a}})(\nabla_{\mathbf{r},\mathbf{v}}\cdot\hat{\mathbf{J}})P&=(\nabla_{\mathbf{r},\mathbf{v}}\cdot\hat{\mathbf{J}})J_{r_{a}}^{t}-\sum_{i}\delta_{ia}(J_{v_{a}}-\gamma(\beta m_i)^{-1}\partial_{v_{i}}P)  \nonumber \\ &\hspace{2cm}+ \sum_{i}(\beta m_i)^{-1}\delta_{ia}(\partial_{r_{i}}+\gamma \partial_{v_{i}})P\\
(\hat{J}_{v_{a}}+\hat{T}_{v_{a}})(\nabla_{\mathbf{r},\mathbf{v}}\cdot \hat{\mathbf{J}})P &=(\nabla_{\mathbf{r},\mathbf{v}}\cdot\hat{\mathbf{J}})J_{v_{a}}^{t}+\sum_{i}\delta_{ia}(\gamma J_{v_{a}}-\frac{\gamma}{\beta m_i} \partial_{r_{i}}P+\frac{1}{m_i}\frac{\partial^{2}V}{\partial r_{i}\partial r_{a}}J_{r_{i}}) \nonumber \\ 
& \hspace{2cm} +\frac{1}{\beta {m_i}^2} \frac{\partial^{2}V}{\partial r_{i} \partial r_{a}}\partial_{v_{i}}P 
\end{align}\end{subequations} that can be recasted as 
\begin{equation}
\mathbf{\hat{\mathbf{J^t}}(\nabla_{\mathbf{r},\mathbf{v}}\cdot\hat{\mathbf{J}})}P=(\nabla_{\mathbf{r},\mathbf{v}}\cdot\hat{\mathbf{J}})\hat{\mathbf{J^t}}P+\left(\begin{array}{cc}
0 & -\delta_{ia}\\
\frac{1}{m_i}\frac{\partial^{2}V}{\partial r_{i}\partial r_{a}} & \gamma\delta_{ia}\end{array}\right)\left(\begin{array}{c}
J_{r_{a}}+(\beta m_i)^{-1}\partial_{v_{a}}P\\
J_{v_{a}}-(\beta m_i)^{-1}\partial_{r_{a}}P\end{array}\right)\end{equation}
leading to equation \eqref{eqn:transcurdyn}.

\section{Appendix B}\label{app:ex1d}
In our study of the LJ38 system, we carried out simulations of the
stochastic dynamics of the clones but did not explicitly make the
averages~\eqref{eqn:vectaverage} that would yield the transition
current. We argued in section~\ref{Sec:evolcuralgo} that if one only
looks for the reaction paths, then these averages are not necessary.
In this appendix, we illustrate this claim on a simple 1d example that
also allows us to discuss several aspects of the clone  dynamics,
namely the metastability, the finiteness of the clone population and
the role of the initial condition.

\begin{figure}[b]
  \begin{center}
\includegraphics{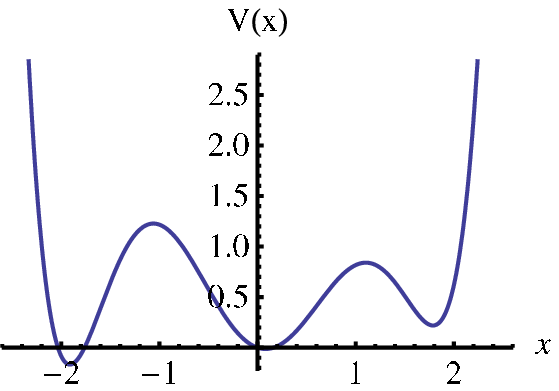} \raisebox{-.6cm}{\includegraphics{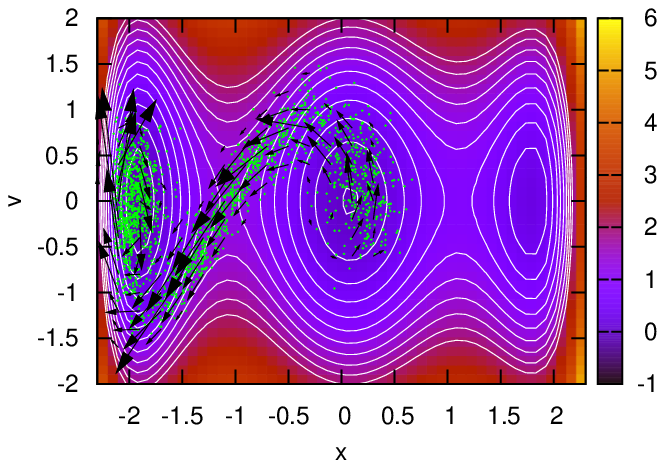}}
  \end{center}
  \caption{{\bf Left}: Plot of the potential $V(x)=x (-39 +240 x + 15
    x^2 -138 x^3+20 x^5)/120$. {\bf Right}: The green crosses are the
    position of the 2000 clones after a time $t=400$. The black arrows
    correspond to the averages~\eqref{eqn:appav} and indeed
    point tangentially to the reaction path. The color code and
    contour lines corresponds to the value of the Hamiltonian
    $H(x,v)=v^2/2+V(x)$}
  \label{fig:potentiel}
\end{figure}

We thus consider a system undergoing an underdamped Langevin dynamics
\begin{equation}
  \dot x=v;\qquad \dot v = -\gamma v - m^{-1} V'(x)+\sqrt{2 \gamma k T/m}\, \eta
\end{equation}
where $V(x)$ is a potential with two barriers, plotted in
figure~\ref{fig:potentiel}. We ran the clone dynamics for 2000 clones
starting in the left well and carried out the
averages~\eqref{eqn:vectaverage}.

To do so, we constructed an approximate density from the positions and
vectors $(x^i,v^i,u_x^i,u_v^i)$ of each of the ${\cal N}_c=2000$ clones:
\begin{equation}
  \label{eqn:fnum}
  {\cal F}_{\rm num}(x,v,u_x,u_v)=\frac 1 {{\cal N}_c}\sum_{i=1}^{{\cal N}_c} \delta(x-x^i) \delta(v-v^i) \delta(u_x-u_{x}^i)\delta(u_v-u_{v}^i)
\end{equation}
In principle, the $\delta$ should be Dirac functions but for practical
purposes we replaced the one acting on the phase space coordinate by the bell-shaped function
\begin{equation}
  \label{eqn:dnum}
  \delta_n(x,v)=\frac 1 Z \exp\left(-\frac 1
        {1-\frac{x^2+v^2}{w^2}}\right) \quad
        \text{if}\quad x^2+v^2<w^2\quad\text{and}\quad   \delta_n(x,v)=0 \quad \text{otherwise}
\end{equation}
where $w=0.1$ and $Z$ is a normalization constant. Finally,
using~\eqref{eqn:fnum} and \eqref{eqn:dnum}
in~\eqref{eqn:vectaverage}, we construct the transition current from
the numerical data by computing
\begin{equation}
  \label{eqn:appav}
  {\bf J^T}(x,v)=\frac{1}{{\cal N}_c}\sum_{i=1}^{{\cal N}_c} {\bf u}^i
  \, \delta_n(x-x_i,v-v_i)
\end{equation}
on a grid every $dx=dy=0.15$ and plot the resulting vector if its norm
is larger than $10^{-3}$. For visualization purposes, we plot in the
figures the vectors $5$ times longer than they really are.

We started a simulation with $2000$ clones in the left well, around
$x=m_1\simeq-1.9$, with unitary vectors $(u_x,u_y)$ pointing at random. The
temperature is set to $kT=0.09$ and the friction to $\gamma=1.5$ so
that the mean first passage time across the barrier is~\cite{HTB1990}
\begin{equation}
  T_{L\to C} \simeq \frac{2\pi } {
    \sqrt{\gamma^2/4 + |V''(M_1)|} - \gamma/2} \sqrt{\frac{|V''(M_1)|}{V''(m_1)}}  e^{\frac{V(M_1)-V(m_1)}{kT}} \simeq  10^7
\end{equation}
where $M_1$ is the first maximum of the potential $M_1\simeq -1$. The
results of the simulation after a time $t=400$ are plotted in
figure~\ref{fig:potentiel}. The clones have already populated the
barrier.~\cite{foot5} As can be seen, the
averages~\eqref{eqn:appav} along the reaction path are non-zero and
result in vectors tangent to the reaction path, pointing toward the
left well.

At later times, two processes take place, roughly on the same time
scale. Firstly, more and more clones come back from the central well
to the left one. Their vectors ${\bf u}$ are always tangent to their
trajectories, but can be pointing toward the left or the central well
with equal probability. Indeed, if $(p(t),q(t),{\bf u}(t))$ is a
possible trajectory of the system, then so is $(p(t),q(t),-{\bf
  u}(t))$. As a result, the averages~\eqref{eqn:appav} may cancel out
at large times, when the subpopulations of clones whose vectors ${\bf
  u}$ point toward the central and the left wells balance. This is how
numerically the transition current is supposed to vanish at large time
(another possibility being that all the clones leave a region of
phase-space, because of finite population-size effects).

\begin{figure}[h!]
  \begin{center}
\raisebox{-.6cm}{\includegraphics{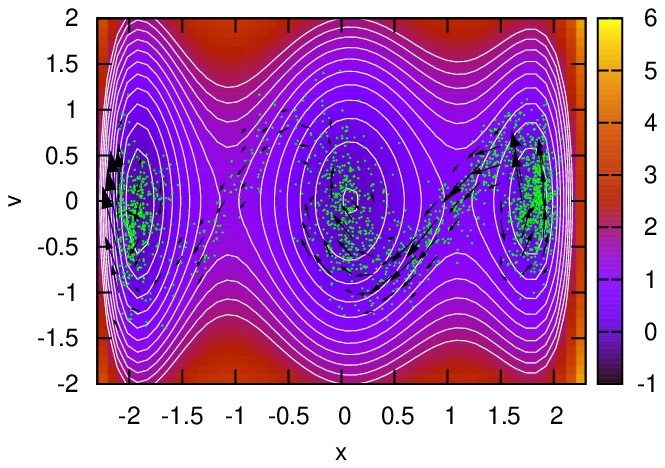}}\hspace{1cm}  \raisebox{-.6cm}{\includegraphics{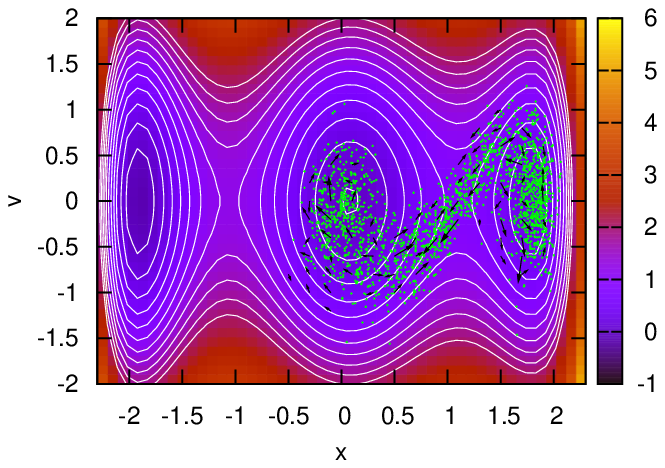}}
  \end{center}
  \caption{{\bf Left}: $t=997$ the clones populate both barriers. The
    arrows average out along the reaction path between the left and
    central wells which have equilibrated, whereas the transition
    current is still present between the central and right wells {\bf
      Right}: at $t=1590$ the clones only populate the reaction path
    between the central and right wells. Since the simulation had
    enough time to equilibrate the involved wells, the
    average~\eqref{eqn:vectaverage} cancels out.}
  \label{fig:clones1dlater}
\end{figure}

Secondly, some clones reach the barrier leading to the right well and
duplicate, which results in populating the second reaction path. Since
the clones did not have time to fall in the right well and cross back
the barrier towards the central well with vectors ${\bf u}$ pointing
in the opposite direction, the average~\eqref{eqn:appav} does not cancel
along this reaction path.

Both effects can be seen in figure~\ref{fig:clones1dlater}: the clones
populate both barriers; the average~\eqref{eqn:appav} cancels out over
the first barrier but not yet over the second one. This shows that
the clone dynamics do locate the barriers and remain on the reaction
paths even though the transition current may average out when the two
wells separating a barrier equilibrate. This thus validates our
approach to locate the reaction paths in the LJ38 system.

Note that if one wishes to study quantitatively the transition
current, two modifications would need to be done to our
algorithm. First, the initial condition should not be taken at random
but constructed as proposed in section~\ref{sec:fp}. Second, rather
than simulating all the clones in parallel while maintaining their
population constant, it may be advantageous to run them sequentially,
starting one run for every offspring of every clones, as is done for
instance with the `Go with the winner' methods.~\cite{G2002} The
constrain on the total population being fixed indeed affects the
metastability of the clone dynamics and increases finite size
effects. For instance, if there are ${\cal N}_1$ and ${ \cal N}_2$
clones on the same reaction path, with vectors pointing in opposite
directions, both populations grow exponentially with the same rate.
Now, if the total population is kept constant, then the smallest
sub-population disappears on average exponentially fast. Using a `Go
with the winner' method would palliate this drawback but would result
in additional computational costs difficult to estimate beforehand.

\bibliographystyle{elsarticle-num}

\begin{thebibliography}{99}
\bibitem{foot4}{The vanishing of the current on a given branch
   appears usually in the program as a depopulation of clones in that
   region, but it can also be that their vectors are in counter-phase,
   so that their contribution to the current vanishes.}
\bibitem{foot5} Note that with standard Langevin simulations of the
  same duration, the probability that {\em none} of the 2000 clones
  has crossed the barrier is more than 90\%.
\end{thebibliography}

\end{document}